\documentclass{article}
\usepackage{bm}
\begin{document}

\title{Neutrino masses from an approximate mixing matrix with $\theta_{13}\neq 0$} 
\author{Asan Damanik\footnote{E-mal: d.asan@lycos.com}\\ {\it Department of Physics Education, Sanata Dharma University}\\ {\it Kampus III USD Paingan Maguwoharjo Sleman Yogyakarta, Indonesia}}
\date{}

\maketitle

\abstract{An approximate neutrino mixing matrix is formutated by using the standard neutrino mixing matrix as a basis and experimental data of neutrino oscillations as inputs.  By using the resulted approximate neutrino mixing matrix to proceed the neutrino mass matrix and constraining the resulted neutrino mass matrix with zero texture: $M_{\nu}(1,1)=M_{\nu}(1,3)=M_{\nu}(3,1)=0$, we can have neutrino masses as function of mixing angle $\theta_{13}$ with normal hierarchy: $m_{1}<m_{2}<m_{3}$.  By taking the central value of mixing angle $\theta_{13}=9^{o}$ that gives $\epsilon=0.16$ and using the squared mass difference: $\Delta m_{32}^{2}$ for normal hierarchy, we then obtained neutrino masses: $m_{1}=0.00847$ eV, $m_{2}=0.01215$ eV, and $m_{3}=0.05062$ eV which can predict the squared mass difference for solar neutrino precisely with the experimental result: $\Delta m_{21}^{2}=7.59\times 10^{-5}~{\rm eV^{2}}$.}

\section{Introduction}
Since the Kamiokande Collaboration reported the evidence of neutrino oscillations \cite{Fukuda}, the neutrino oscillation concept, which is previously proposed in order to explain the solar neutrino deficit, have been accepted by physicists as a new elegant concept in leptonic sector especially in neutrino sector which have many consequences for the new physics beyond the standard model.   The theorists have made some attempts to explain the neutrino flavor oscillation.  One of the most elegant idea to explain the neutrino oscillation is adopting the concept that neutrino flavor eigenstate is not its own mass eigenstate, but neutrino flavor eigenstates ($\nu_{e},\nu_{\mu},\nu_{\tau}$ ) are linear combination of neutrino mass eigenstates ($\nu_{1},\nu_{2},\nu_{3}$ ) as follow:
\begin{eqnarray}
\nu_{\alpha}=V_{\alpha\beta}\nu_{\beta},\label{1}
\end{eqnarray}
where the indexes $\alpha=e,\mu,\tau$, $\beta=1,2,3$ , and $V_{\alpha\beta}$ are the elements of the neutrino mixing $V$.  The  standard parameterization of the mixing matrix ($V$) read:
\begin{eqnarray}
V=\bordermatrix{& & &\cr
&c_{12}c_{13} &s_{12}c_{13} &s_{13}e^{-i\delta}\cr
&-s_{12}c_{23}-c_{12}s_{23}s_{13}e^{i\delta} &c_{12}c_{23}-s_{12} s_{23}s_{13}e^{i\delta}&s_{23}c_{13}\cr
&s_{12}s_{23}-c_{12}c_{23}s_{13}e^{i\delta} &-c_{12}s_{23}-s_{12}c_{23}s_{13}e^{i\delta} &c_{23}c_{13}}
 \label{V}
\end{eqnarray}
where $c_{ij}$ is the $\cos\theta_{ij}$, $s_{ij}$ is the $\sin\theta_{ij}$, $\theta_{ij}$ are the mixing angles, and $\delta$ is the Dirac CP-violating phase.

It is apparent from Eq. (\ref{V}) that from the beginning of the formulation of neutrino mixing matrix, the nonzero of the mixing angle $\theta_{13}$ have been anticipated together with the Dirac phase as fundamental parameters in neutrino mixing matrix.  For many years before the reported experimental results by T2K \cite{T2K}, MINOS \cite{MINOS}, Double Chooz \cite{Double}, Daya Bay \cite {Daya}, and RENO \cite{RENO}, the mixing angle $\theta_{13}$ to be put in zero or approximately zero due to the experiments limitations to measure it precisely.  We have three well-known neutrino mixing matrices with mixing angle $\theta_{13}=0$, i.e. bimaximal mixing (BM), tribimaximal mixing (TBM), and democratic mixing (DM).  Recently, the experimental results showed that the mixing angle $\theta_{13}\neq 0$ and relatively large which imply the three well-known neutrino mixing matrices should be ruled out or modified.

From theoretical side, several theoretical attempt have already been performed by many authors in order to accommodate the nonzero and relatively large mixing angle $\theta_{13}$  by modifying the neutrino mixing matrix including the Dirac phase $\delta$ in relation to the CP-violation in neutrino sector.  Another unsolved problem in neutrino physics till today is the hierarchy of neutrino mass.  Experimental results showed that we have two possibilities for neutrino mass hierarchies: normal and inverted hierarchies.  We have no clue in order to decide theoretically the neutrino mass hierarchy whether it normal or inverted.

In this paper, we study sistematically and formulate neutrino mixing matrix by using the standard neutrino mixing matrix in Eq. (\ref{V}) by considering the reported neutrino oscillations experimental data as inputs.  The resulted neutrino mixing matrix to be used to obtain the nonzero mixing angle $\theta_{13}$ and its relation to neutrino masses.  In section 2, we show  the simple basic motivations and assumptions to formulate the three well-known mixing matrices: BM, TBM, and DC which cannot proceed nonzero $\theta_{13}$.  In section 3, we use recent neutrino experimental data as input in order to obtain an approximate neutrino mixing matrix and discuss its predictions on neutrino masses. The formulation of our neutrino mixing matrix is  similar with the formulation of BM, TBM, and DC, but we anticipate the nonzero mixing angle: $\theta_{13}$ in our scenario.  Finally, section 4 is devoted to conclusions.

\section{Neutrino mixing marix: BM, TBM, and DC}
In order to give us a concise knowledge and view of the early patterns of neutrino mixing matrices before the era of nonzero mixing angle $\theta_{13}$ which is also known as the reactor mixing angle, in this section we review the three well-known mixing matrices which put mixing angle $\theta_{13}=0$ or approximately zero.  The experimental facts from neutrino oscillation showed us that both solar neutrino mixing angle ($\theta_{12}$) and atmospheric neutrino mixing angle ($\theta_{23}$)  nearly maximal: $\theta_{12}\approx\theta_{23}\approx \pi/4$, and mixing angle: $\theta_{13}\approx 0$.  Thus, for the first approximation, we can write the neutrino mixing in Eq. (\ref{V}) as follow \cite{Barger}-\cite{HeXu}:
\begin{eqnarray}
V_{BM}=\bordermatrix{& & &\cr
&\sqrt{1/2} &\sqrt{1/2} &0\cr
&-1/2 &1/2 &\sqrt{1/2}\cr
&1/2 &-1/2 &\sqrt{1/2}}. \label{BM}
\end{eqnarray}
which is known as bimaximal mixing matrix (BM).  The tribimaximal mixing (TBM) also formulated in accordance with the experimental results of neutrino oscillation: $\theta_{13}\approx 0$ and unitarity constraints.  The first formulation of TBM-like based on experimental facts was formulated by Stancu and Ahluwalia as follow \cite{Stancu}:
\begin{eqnarray}
V_{TBM-like}=\bordermatrix{& & &\cr
&c_{\theta} &s_{\theta} &0\cr
&-s_{\theta}/\sqrt{2} &c_{\theta}/\sqrt{2} &1/\sqrt{2} \cr
&s_{\theta}/\sqrt{2} &-c_{\theta}/\sqrt{2} &1/\sqrt{2}},
\end{eqnarray}
where: $c_{\theta}=\cos\theta$ and $c_{\theta}=\sin\theta$ with $\theta$  is the solar neutrino mixing angle.

Theoretical derivation of TBM can also be read in \cite{Harrison}-\cite{He} as follow:
\begin{eqnarray}
V_{TBM}=\bordermatrix{& & &\cr
&\sqrt{2/3} &\sqrt{1/3} &0\cr
&-\sqrt{1/6} &\sqrt{1/3} &-\sqrt{1/2}\cr
&-\sqrt{1/6} &\sqrt{1/3} &\sqrt{1/2}}. \label{TBM}
\end{eqnarray}
and the democratic mixing also with the approximation: $\theta_{13}\approx 0$ .  The democratic mixing (DC) read \cite{Fritzsch96}-\cite{Fritzschb}:
\begin{eqnarray}
V_{DC}=\bordermatrix{& & &\cr
&\sqrt{1/2} &\sqrt{1/2} &0\cr
&\sqrt{1/6} &-\sqrt{1/6} &-\sqrt{2/3}\cr
&-\sqrt{1/3} &\sqrt{1/3} &-\sqrt{1/3}}. \label{DC}
\end{eqnarray}

From Eqs. (\ref{BM})-(\ref{DC}), one can see that three well-known mixing matrices predict mixing angle $\theta_{13}= 0$ which is incompatible with the present data of neutrino oscillations.  The latest results from neutrino oscillation experiment reported by T2K Collaboration \cite{T2K}:
\begin{eqnarray}
5^{0}\leq\theta_{13}\leq 16^{0},\label{NH}
\end{eqnarray}
for neutrino mass in normal hierarchy (NH), and
\begin{eqnarray}
5.8^{0}\leq\theta_{13}\leq 17.8^{0},\label{IH}
\end{eqnarray}                
for inverted hierarchy (IH) with Dirac phase: $\delta=0$.  The nonzero value of mixing angle: $\theta_{13}$   was also confirmed by Daya Bay Collaboration \cite{Daya}] as follow:
\begin{eqnarray}
\sin^{2}2\theta_{13}=0.092\pm 0.016 (\rm{stat.})\pm 0.005 (\rm{syst.}).\label{D}
\end{eqnarray}
Thus, we should modify or reexamine the three well-known neutrino mixing matrices which are shown in Eqs. (\ref{BM})-(\ref{DC}) because it cannot anymore accommodate the present data precisely.  The modified neutrino mixing matrix which can accommodate the nonzero and relatively large mixing angle $\theta_{13}$ can be read in Refs. \cite{Hea}-\cite{Damanik2}.

\section{Neutrino mass from an approximate mixing matrix}
As stated in section 1, this section is devoted to formulate an approximate neutrino mixing matrix based on standard parametrization by considering the experimental data from neutrino oscillation experiments as inputs.  Recently, it is apparent that the mixing angle $\theta_{13}$  is nonzero and relatively large as shown in Eqs. (\ref{NH}), (\ref{IH}), and (\ref{D}). By using the standard neutrino mixing matrix in  Eq. (\ref{V}) as a basis in formulating an approximate neutrino mixing matrix and experimental facts that mixing angle: $\theta_{13}\neq 0$ and we replace $s_{13}$ as follow :
\begin{eqnarray}
s_{13}=\sin\theta_{13}=\epsilon,
\end{eqnarray}
and it implies:
\begin{eqnarray}
c_{13}=\sqrt{1-\epsilon^{2}},
\end{eqnarray}
then the neutrino mixing matrix in Eq. (\ref{V}) read:
\begin{eqnarray}
V_{A}=\bordermatrix{& & &\cr
&c_{12}\sqrt{1-\epsilon^{2}} &s_{12}\sqrt{1-\epsilon^{2}} &\epsilon\cr
&-s_{12}c_{23}-c_{12}s_{23}\epsilon &c_{12}c_{23}-s_{12}s_{23}\epsilon &s_{23}\sqrt{1-\epsilon^{2}} \cr
&s_{12}s_{23}-c_{12}c_{23}\epsilon &-c_{12}s_{23}-s_{12}c_{23}\epsilon &c_{23}\sqrt{1-\epsilon^{2}} }. \label{VA}
\end{eqnarray}
for $\delta=0$. 

By putting mixing angle: $\theta_{23}\approx\pi/4$ (atmospheric neutrino mixing angle is nearly maximal) which it implies: $c_{23}=s_{23}\approx\frac{\sqrt{2}}{2}$, then the approximate mixing angle in Eq. (\ref{VA}) can be written as follow:
\begin{eqnarray}
V_{A}=\bordermatrix{& & &\cr
&c_{12}\sqrt{1-\epsilon^{2}} &s_{12}\sqrt{1-\epsilon^{2}} &\epsilon\cr
&-\frac{\sqrt{2}}{2}(s_{12}+c_{12}\epsilon) &\frac{\sqrt{2}}{2}(c_{12}-s_{12}\epsilon) &\frac{\sqrt{2}}{2}\sqrt{1-\epsilon^{2}} \cr
&\frac{\sqrt{2}}{2}(s_{12}-c_{12}\epsilon) &-\frac{\sqrt{2}}{2}(c_{12}+s_{12}\epsilon) &\frac{\sqrt{2}}{2}\sqrt{1-\epsilon^{2}} }. \label{VA1}
\end{eqnarray}

As dictated by experimental results on solar neutrino mixing angle: $\theta_{12}\approx 35^{\rm o}$, then we can take the approximation for the values of $s_{12}$ and $c_{12}$ as follow:
\begin{eqnarray}
s_{12}\approx\frac{\sqrt{3}}{3}~~{\rm and }~~c_{12}\approx\frac{\sqrt{6}}{3}. \label{c12}
\end{eqnarray}
If we insert the values of $s_{12}$ and $c_{12}$ in Eq. (\ref{c12}) into approximate neutrino mixing matrix in Eq. (\ref{VA1}), then we have:
\begin{eqnarray}
V_{A}=\bordermatrix{& & &\cr
&\frac{\sqrt{6(1-\epsilon^{2})}}{3}  &\frac{\sqrt{3(1-\epsilon^{2})}}{3} &\epsilon\cr
&-\frac{\sqrt{6}+\sqrt{12}\epsilon}{6} &\frac{\sqrt{12}-\sqrt{6}\epsilon}{6} &\frac{\sqrt{2(1-\epsilon^{2})}}{2} \cr
&\frac{\sqrt{6}-\sqrt{12}\epsilon}{6} &-\frac{\sqrt{12}+\sqrt{6}\epsilon}{6} &\frac{\sqrt{2(1-\epsilon^{2})}}{2}}. \label{VA2}
\end{eqnarray}

Now, we are in position to determine the neutrino mixing by using the approximate neutrino mixing matrix $V_{A}$ as shown in Eq. (\ref{VA2}).  If neutrino is a Majorana particle, then neutrino mass matrix ($M_{\nu}$) can be obtained via the following equation:
\begin{eqnarray}
M_{\nu}=VMV^{T},\label{3}
\end{eqnarray}
where $M$ is the neutrino mass matrix in mass basis and $V$ is the mixing matrix.  By using approximate neutrino mixing matrix in Eq. (\ref{VA2}) and neutrino mass matrix in mass basis as follow :
\begin{eqnarray}
M=\bordermatrix{& & &\cr
&m_{1} &0 &0\cr
&0 &m_{2} &0\cr
&0 &0 &m_{3}}, \label{M}
\end{eqnarray}
then we have neitrino mass matrix with pattern:
\begin{eqnarray}
M_{\nu}=\bordermatrix{& & &\cr
&A &B &C\cr
&B &D &E \cr
&C &E &F}, \label{Mnu}
\end{eqnarray}
where:
\begin{eqnarray}
A=\frac{2}{3}(1-\epsilon^{2})m_{1}+\frac{1}{3}(1-\epsilon^{2})m_{2}+\epsilon^{2}m_{3},\\
B=-\frac{\sqrt{1-\epsilon^{2}}}{3}(\sqrt{2}\epsilon+1)m_{1}+\frac{\sqrt{1-\epsilon^{2}}}{6}(2-\frac{\sqrt{2}}{2}\epsilon)m_{2}+\frac{\sqrt{2-2\epsilon^{2}}}{2}\epsilon m_{3},\\
C=-\frac{\sqrt{1-\epsilon^{2}}}{3}(\sqrt{2}\epsilon-1)m_{1}+\frac{\sqrt{1-\epsilon^{2}}}{6}(-2-\frac{\sqrt{2}}{2}\epsilon)m_{2}+\frac{\sqrt{2-2\epsilon^{2}}}{2}\epsilon m_{3},\\
D=\frac{1}{3}(\frac{1}{2}+\sqrt{2}\epsilon+\epsilon^{2})m_{1}+\frac{1}{3}(1-\sqrt{2}\epsilon+\frac{1}{2}\epsilon^{2})m_{2}+\frac{1}{2}(1-\epsilon^{2})m_{3},\\
E=-\frac{1}{3}(\frac{1}{2}-\epsilon^{2})m_{1}-\frac{1}{3}(1-\frac{\epsilon^{2}}{2}m_{2}+\frac{1}{2}(1-\epsilon^{2})m_{3},\\
F=\frac{1}{3}(\frac{1}{2}-\sqrt{2}\epsilon+\epsilon^{2})m_{1}+\frac{1}{3}(1-\sqrt{2}\epsilon+\frac{1}{2}\epsilon^{2})m_{2}+\frac{1}{2}(1-\epsilon^{2})m_{3}.
\end{eqnarray}

By inspecting the entries of resulted neutrino mass matrix ($M_{\nu}$), it is apparent that all entries of  $M_{\nu}$ relate with $\epsilon$.  In order to find the relation of $\epsilon$ with the measured parameters in neutrino oscillation experiments, we simplify the problem by imposing texture zero into neutrino mass matrix in Eq. (\ref{Mnu}).  The texture zero analysis is also known as Fritzsch anzatze that can be used to solve the problem of quark-lepton mass matrices \cite{Fritzschtz}.  The texture zero can help us to simplify the problem becauseit reduce the number of parameters or degre of fredom in neutrino mass matrix.  If we impose the zero texture into resulted neutrino mass matrix in Eq. (\ref{Mnu}) i.e. $A=M_{\nu}(1,1)=0$ and $C=M_{\nu}(1,3)=0$, then we have the following relations:
\begin{eqnarray}
m_{1}=m_{2}-\frac{3\sqrt{2}}{2}\epsilon m_{3},\label{m1}
\end{eqnarray}
and
\begin{eqnarray}
m_{1}=\frac{1}{2}m_{2}-\frac{3}{2}\frac{\epsilon^{2}}{\epsilon^{2}-1}m_{3}, \label{m11}
\end{eqnarray}
respectively.  From Eqs. (\ref{m1}) and (\ref{m11}) we can have the relation:
\begin{eqnarray}
\frac{m_{2}}{m_{3}}=\sqrt{2}\epsilon+\frac{\epsilon^{2}}{\epsilon^{2}-1},\label{e}
\end{eqnarray}
and finally we can also rewrite Eq. (\ref{m1}) or (\ref{m11}) as follow:
\begin{eqnarray}
\frac{m_{1}}{m_{2}}=\frac{1}{2}-\frac{3}{2}\frac{\epsilon}{\epsilon+\sqrt{2}(\epsilon^{2}-1)}. \label{m12}
\end{eqnarray}
The texture zero pattern similar with our texture zero pattern has also been used for a model of quark and lepton mass matrices in \cite{Nishiura} which give a consistent prediction with the experimental data.  A different viable textures for quarks, where only the Cabibbo mixing arises from the down sector, and extend to the charged leptons while constructing a complementary neutrino structure that leads to viable lepton masses and mixing can be read in \cite{Hernandez}.

By referring to experimental value of mixing angle $\theta_{13}$ in Eq. (\ref{NH}), the value of $\epsilon$ reads:
\begin{eqnarray}
0.087\leq\epsilon\leq 0.276,
\end{eqnarray}
which proceed:
\begin{eqnarray}
\frac{m_{1}}{m_{2}}<1,~\frac{m_{2}}{m_{3}}<1,\label{H}
\end{eqnarray}
It is apparent from Eq. (\ref{H}) that approximate neutrino mixing matrix  predict neurino masses in normal hierarchy: $m_{1}< m_{2}< m_{3}$.

From Eq. (\ref{e}) we can have the squared mass difference of atmospheric neutrino as follow:
\begin{eqnarray}
\Delta m_{32}^{2}=m_{3}^{2}-m_{2}^{2}=m_{3}^{2}\left[1-\left(\sqrt{2}\epsilon+\frac{\epsilon^{2}}{\epsilon^{2}-1}\right)^{2}\right],\label{smd}
\end{eqnarray}
which proceed:
\begin{eqnarray}
m_{3}=\sqrt{\frac{\Delta m_{32}^{2}}{1-\left(\sqrt{2}\epsilon+\frac{\epsilon^{2}}{\epsilon^{2}-1}\right)^{2}}}.\label{M3}
\end{eqnarray}
By the same procedure in obtaining squared mass difference for atmospheric neutrino, the squared mass difference for solar neutrino can be derived from Eq. (\ref{m12}) which proceed:
\begin{eqnarray}
\Delta m_{21}^{2}=m_{2}^{2}-m_{1}^{2}=m_{2}^{2}\left[1-\left(\frac{1}{2}-\frac{3}{2}\frac{\epsilon}{\epsilon+\sqrt{2}(\epsilon^{2}-1)}\right)^{2}\right],\label{smd1}
\end{eqnarray}
which give:
\begin{eqnarray}
m_{2}=\sqrt{\frac{\Delta m_{21}^{2}}{1-\left(\frac{1}{2}-\frac{3}{2}\frac{\epsilon}{\epsilon+\sqrt{2}(\epsilon^{2}-1)}\right)^{2}}}.\label{M2}
\end{eqnarray}

In order to determine neutrino masses, we take the experimental data of neutrino oscillation as input.  The current combined world data of neutrino oscillations especially squared mass difference read \cite{Gonzales,Fogli}:
\begin{eqnarray}
\Delta m_{21}^{2}=7.59\pm 0.20(^{+0.61}_{-0.69})\times 10^{-5}~{\rm eV^{2}},~~~~~~~~~\label{21}\\
\Delta m_{32}^{2}=2.46\pm 0.12(\pm 0.37)\times 10^{-3}~{\rm eV^{2}}, ~{\rm for~ NH}, \label{32N}\\
\Delta m_{32}^{2}=-2.36\pm 0.11(\pm 0.37)\times 10^{-3}~{\rm eV^{2}}. ~{\rm for~ IH,} \label{32I}
\end{eqnarray}

If we insert the central value of $\epsilon=0.16$ into Eqs. (\ref{M3}) and (\ref{M2}), and the squared mass difference: $\Delta m_{32}^{2}$ for normal hierarchy as shown in Eq. (\ref{32N}), then we can determine neutrino masses $m_{2}$ and $m_{3}$.  By knowing the neutrino masses $m_{2}$ and $m_{3}$, we can determine the neutrino mass $m_{1}$ from Eq. (\ref{m1}) or (\ref{m11}).   The obtained neutrino masses whithin this scenario read:
\begin{eqnarray}
m_{1}=0.00847~{\rm eV},\\
m_{2}=0.01215~{\rm eV},\\
m_{3}=0.05062~{\rm eV},
\end{eqnarray}
which can predict the squared mass difference for solar neutrino precisely with the experimental result:
\begin{eqnarray}
\Delta m_{21}^{2}=7.59\times 10^{-5}~{\rm eV^{2}}.
\end{eqnarray}

\section{Conclusions}
We have formulated and studied systematically the approximate neutrino mixing matrix by using the standard neutrino mixing matrix as a basis and considering the experimental data of neutrino oscillations as inputs.  By using the resulted approximate neutrino mixing matrix to proceed the neutrino mass matrix and constraining the resulted neutrino mass matrix with zero texture: $M_{\nu}(1,1)=M_{\nu}(1,3)=M_{\nu}(3,1)=0$, we can have neutrino masses as function of mixing angle $\theta_{13}$ with normal hierarchy: $m_{1}<m_{2}<m_{3}$.   By taking the central value of mixing angle $\theta_{13}=9^{o}$ that gives $\epsilon=0.16$ and using the squared mass difference: $\Delta m_{32}^{2}$ for normal hierarchy, we then obtained neutrino masses: $m_{1}=0.00847$ eV, $m_{2}=0.01215$ eV, and $m_{3}=0.05062$ eV which can predict the squared mass difference for solar neutrino precisely with the experimental result: $\Delta m_{21}^{2}=7.59\times 10^{-5}~{\rm eV^{2}}$.


\begin{thebibliography}{00}
\bibitem{Fukuda}
Kamiokande Collab.(Y. Fukuda {\it et al.}), Evidence for oscillation of atmospheric neutrinos, {\it Phys. Rev. Lett.} {\bf 81} (1998), 1562-1567.
\bibitem{T2K}
T2K Collab. (K. Abe {\it et al.}), Indication of electron neutrino apperance from an accelerator produced off-axis muon neutrino beam,  {\it Phys. Rev. Lett.} {\bf 107} (2011), 041801.
\bibitem{MINOS}
MINOs Collab. (P. Adamson {\it et al.}), Improved search for muon-neutrino to electron-neutrino oscillations in MINOS, {\it Phys. Rev. Lett.} {\bf 107} (2011), 181802.
\bibitem{Double}
Double-Chooz Collab. (Y. Abe {\it et al.}), Indication of reactor $\overline{\nu}_{e}$ disappearance in the Double Chooz experiment, {\it Phys. Rev. Lett.} {\bf 108} (2012), 131801.
\bibitem{Daya}
Daya Bay Collab. (F. P. An {\it et al.}), Observation of electron-antineutrino disappearance at Daya Bay, {\it Phys. Rev. Lett.} {\bf 108 (2012)}, 171803.
\bibitem{RENO}
RENO Collab. (J. K. Ahn {\it et al.}), Observation of electron antineutrinos disappearance in the RENO experiment, {\it Phys. Rev. Lett.} {\bf 108} (2012), 191802.
\bibitem{Barger}
V. D. Barger, S. Pakvasa, T. J. Weiler, and K. Whisnant, Bi-maximal oscillation solutions for solar and atmospheric neutrinos, {\it Phys. Lett.} {\bf B 437} (1998), 107-116.
\bibitem{Baltz}
A. J. Baltz, A. S. Goldhaber, and M. Goldhaber, Solar Neutrino Puzzle: An Oscillation Solution with Maximal Neutrino Mixing, {\it Phys. Rev. Lett.} {\bf 81} (1998), 5730.
\bibitem{Georgi}
H. Georgi and S. L. Glashow, Neutrinos on earth and in the heaavens, {\it Phys. Rev.} {\bf D 61} (2000), 097301.
\bibitem{HX}
D. A. Dicus, H.-J. He and J. N. Ng, Minimal schemes for large neutrino mixing with inverted hierarchy, {\it Phys. Lett.} {\bf B 363} (2002) 83.
\bibitem{Li}
N. Li and B.-Q. Ma, A new parametrization of the neutrino mixing matrix, {\it Phys. Lett.} {\bf B 600} (2004), 248-254.
\bibitem{HeXu}
H.-J. He and X.-J. Xu, Octahedral Symmetry with geometrical breaking: New prediction for neutrino mixing angle $\theta_{13}$ and CP violation, {\it Phys. Rev.} {\bf D 86} (2012) 111301.
\bibitem{Stancu}
I. Stancu and D. V. Ahluwalia, L/E-flatness of the electron-like event ratio, in Super-Kamiokande and a degeneracy, in neutrino masses, {\it Phys. Lett.} {\bf B 460} (1999), 431-436.
\bibitem{Harrison}
P. F. Harrison, D. H. Perkins, and W. G. Scott, A Redetermination of the Neutrino Mass-Squared Difference in Tri-Maximal Mixing with Terrestrial Matter Effects, {\it Phys. Lett.} {\bf B 458} (1999), 79-92.
\bibitem{Harrisona}
P. F. Harrison, D. H. Perkins, and W. G. Scott, Tri-Bimaximal mixing and the neutrino oscillation data, {\it Phys. Lett.} {\bf B 530} (2002), 167-173.
\bibitem{Xing}
Z-z. Xing, A full determination of the neutrino mass spectrum from two-zero textures of the neutrino mass matrix, {\it Phys. Lett.} {\bf B 533} (2002), 85-90.
\bibitem{Harrisonb}
P. F. Harrison and W. G. Scott, Symmetries and generalisations of Tri-Bimaximal neutrino mixing, {\it Phys. Lett.} {\bf B 535} (2002), 163-169.
\bibitem{Harrisonc}
P. F. Harrison and W. G. Scott, Permutation symmetry, Tri-Bimaximal neutrino mixing and the S3 group characters, {\it Phys. Lett.} {\bf B 557} (2003), 76-86.
\bibitem{He}
X.-G. He and A. Zee, Some simple mixing and mass matrices for neutrinos, {\it Phys. Lett.} {\bf B 560} (2003), 87-90.
\bibitem{Fritzsch96}
H. Fritzsch and Z-z. Xing, Lepton mass hierarchy and neutrino oscillations, {\it Phys. Lett.} {\bf B 372} (1996), 265-270.
\bibitem{Fritzscha}
H. Fritzsch and Z-z. Xing, Large leptonic flavor mixing and the mass spectrum of leptons, {\it Phys. Lett.} {\bf B440} (1998), 313-318.
\bibitem{Fritzschb}
H. Fritzsch and Z-z. Xing, Maximal neutrino mixing and maximal CP violation, {\it Phys. Rev.} {\bf D 61} (2000), 073016.
\bibitem{Hea}
X-G. He and A. Zee, Minimal modification to Tri-bimaximal mixing, {\it Phys. Rev.} {\bf D 84} (2011), 053004, arXiv:1106.4359 [hep-ph].
\bibitem{Shimizu}
Y. Shimizu, M. Tanimoto, and A. Watanabe, Breaking Tri-bimaximal mixing and large $\theta_{13}$, {\it Prog. Theor. Phys.} {\bf 126} (2011), 81-90, arXiv:1105.2929 [hep-ph].
\bibitem{Wchao}
W. Chao and Y. Zheng, Relatively large theta13 from modification to the Tri-bimaximal, Bimaximal and
Democratic neutrino mixing matrices, arXiv:1107.0738 [hep-ph].
\bibitem{Xing11}
Z-z.Xing, T2K indication of relatively large $\theta_{13}$ and a natural perturbation to the democratic neutrino mixing pattern, {\it Chin. Phys.} {\bf C 36} (2012),101-105, arXiv:1106.3244 [hep-ph].
\bibitem{Deepthi11}
K. N. Deepthi, S. Gollu, and R. Mohanta, Neutrino mixing matrices with relatively large $\theta_{13}$ and with texture one-zero, {\it Eur. Phys. J.} {\bf C 72} (2012), 1888, arXiv:1111.2781v1 [hep-ph].
\bibitem{Damanik1}
A. Damanik, Nonzero $\theta_{13} $ and neutrino masses from modified neutrino mixing matrix, {\it Int. J. Mod. Phys.} {\bf A 27} (2012), 1250091.
\bibitem{Damanik2}
A. Damanik, Nonzero $\theta_{13}$ and neutrino masses from modified TBM, {\it EJTP} {\bf 11} (2014) (31), 125-130.\bibitem{Fritzschtz}
H. Fritzsch, Weak interaction mixing in the six-quark theory, {\it Phys. Lett.} {\bf B 73} (1978) 317-322.
\bibitem{Nishiura}
H. Nishiura, K. Matsuda, and T. Fukuyama, Lepton and quark mass matrices, {\it Phys. Rev.} {\bf D 60} (1999) 013006.
\bibitem{Hernandez}
A. E. C. Hernandez and I. de Medeiros Varzielas, Viable textures for the fermion sector, {\it J. Phys.} {\bf G 42} (2015) 065002.
\bibitem{Gonzales}
M. Gonzales-Garcia, M. Maltoni and J. Salvado, Updated global fit to three neutrino mixing: status of the hints of $\theta_{13}>0$, {\it JHEP} {\bf 04} (2010), 056.
\bibitem{Fogli}
G. Fogli {\it et al.}, Probing $\theta_{13}$ with global neutrino data analysis, {\it J. Phys. Con. Ser.} {\bf 203} (2010), 012103.
\end{thebibliography}
\end{document}